\documentclass[final, 3p,twocolumn]{elsarticle}

\usepackage{amssymb}
\usepackage{amsmath}
\usepackage{amsthm}
\usepackage{amsfonts}       
\usepackage{nicefrac}       
\newcommand{\R}{{\mathbb R}} 
\newcommand{\AstroClearNet}{\emph{AstroClearNet}} 

\theoremstyle{remark}

\usepackage[utf8]{inputenc} 
\usepackage[T1]{fontenc}    
\usepackage{hyperref}       
\usepackage{url}            
\usepackage{booktabs}       
\usepackage{multirow}

\usepackage{float}
\usepackage{microtype}      
\usepackage{lipsum}
\usepackage{fancyhdr}       
\usepackage{subcaption}
\usepackage{indentfirst}
\usepackage{xcolor}
\usepackage{mathtools}

\journal{Astronomy and Computing}

\begin{document}

\begin{frontmatter}

\title{AstroClearNet: Deep image prior for multi-frame astronomical image restoration}

\author[ams]{Yashil Sukurdeep\corref{cor}} 
\cortext[cor]{Corresponding author: Yashil Sukurdeep\\
\texttt{yashil.sukurdeep@jhu.edu}}

\author[jhu]{Fausto Navarro} 

\author[ams,cs,physics]{Tam\'{a}s Budav\'{a}ri} 

\affiliation[ams]{organization={Department of Applied Mathematics and Statistics, Johns Hopkins University},
            city={Baltimore},
            postcode={21218}, 
            state={MD},
            country={USA}}

\affiliation[jhu]{organization={Johns Hopkins University},
            city={Baltimore},
            postcode={21218}, 
            state={MD},
            country={USA}}
            
\affiliation[cs]{organization={Department of Computer Science, Johns Hopkins University},
            city={Baltimore},
            postcode={21218}, 
            state={MD},
            country={USA}}

\affiliation[physics]{organization={Department of Physics and Astronomy, Johns Hopkins University},
            city={Baltimore},
            postcode={21218}, 
            state={MD},
            country={USA}}

\begin{abstract}
Recovering high-fidelity images of the night sky from blurred observations is a fundamental problem in astronomy, where traditional methods typically fall short. In ground-based astronomy, combining multiple exposures to enhance signal-to-noise ratios is further complicated by variations in the point-spread function caused by atmospheric turbulence. In this work, we present a self-supervised multi-frame method, based on deep image priors, for denoising, deblurring, and coadding ground-based exposures. Central to our approach is a carefully designed convolutional neural network that integrates information across multiple observations and enforces physically motivated constraints. We demonstrate the method’s potential by processing Hyper Suprime-Cam exposures, yielding promising preliminary results with sharper restored images. 
\end{abstract}

\begin{keyword}
deep generative prior \sep ground-based astronomy  \sep astronomy image processing

\end{keyword}

\end{frontmatter}

\section{Introduction}
\label{sec:intro}
The latest ground-based astronomical surveys, such as the Hyper Suprime-Cam (HSC) survey~\citep{aihara2018hyper} and the upcoming Legacy Survey of Space and Time (LSST) from the Rubin Observatory~\citep{ivezic2019lsst}, are designed to capture exposures of vast portions of the night sky. These surveys rely heavily on high-resolution ground-based telescopes that produce massive datasets, necessitating substantial processing before meaningful scientific analysis can occur. In particular, imaging distant and faint celestial objects as part of these large-scale surveys requires advanced image processing algorithms to maximize the extraction of reliable and useful information.

A key challenge when developing such algorithms lies in addressing atmospheric blur, an unwanted but inevitable consequence of ground-based observations. The process of mitigating this blur, known as deconvolution, is particularly complex due to the high noise levels, wide dynamic range, and various artifacts present in the exposures. Our work tackles the challenge of multi-frame astronomical image restoration, focusing on combining multiple noisy and blurry exposures to produce a sharp, unified latent image of the night sky, with the aim of enhancing signal-to-noise ratios in order to facilitate downstream tasks such as reliable photometric analysis of astronomical sources.

\subsection{Related work}
\label{ssec:related_work}
Past approaches for multi-frame image restoration in the field of astronomy include lucky imaging~\cite{tubbs2003lucky}, coaddition~\cite{annis2014sloan}, maximum likelihood estimation~\cite{schulz1993multiframe, zhulina2006multiframe}, and streaming methods~\cite{harmeling2009online, harmeling2010multiframe, hirsch2011online, lee2017robust, lee2017streaming}. These techniques must carefully balance the effective integration of data from multiple observations, while managing noise levels and blur across exposures.

More recently, several machine learning and deep learning frameworks have achieved viable results in a variety of difficult inverse imaging problems, including image restoration. One such approach involves training a convolutional neural network in a supervised manner to deblur images by performing spatial convolution with a large inverse kernel~\cite{xu2014deep}. Other methods use neural networks to enhance or post-process the outputs of classical image restoration methods, such as Wiener or Richardson-Lucy deconvolution, which rely on undoing the blur in the Fourier domain~\cite{schuler2013machine, son2017fast}. These traditional methods also serve as inspiration for other end-to-end deep learning image restoration techniques~\cite{dong2020deep}.

All of these supervised approaches rely on collections of labeled training data, which is a major impediment for their adoption in the field of astronomy, where ground truth observations are costly to acquire. This motivates the need for unsupervised deep learning approaches for astronomical image restoration.

Prominently, the method of \emph{deep image priors}, a self-supervised technique that was introduced by Ulyanov et al.~\cite{ulyanov2018deep}, represents one such avenue. As part of this approach, a so-called ``hourglass'' image generator network is used to parametrize the mapping from a randomly initialized vector to a single degraded image. This mapping can be leveraged to perform a variety of image processing tasks, such as denoising, in-painting, artifact removal, and super-resolution, among others. The framework has been extended to perform blind image deconvolution with natural images~\citep{ren2020neural,asim2020blind}. 

Yet, unsupervised and self-supervised deep learning techniques have yet to fully gain traction in the astronomy community. Indeed, while a few such methods have emerged, such as frameworks for astronomical inverse problems~\cite{lanusse2019hybrid} and the analysis of galaxy images and spectra~\cite{parker2024astroclip}, their potential is yet to be fully unlocked in the domain of multi-frame image restoration, and especially in the context of processing ground-based observational data, where challenges such as atmospheric distortion and variable seeing conditions persist.

\subsection{Contributions}
\label{ssec:contributions}
Building on the aforementioned advances in deep learning for image restoration---and drawing particular inspiration from deep image priors---we present \AstroClearNet: a framework that, to the best of our knowledge, constitutes one of the first self-supervised deep learning approaches for multi-frame image restoration in astronomy.

As part of \AstroClearNet, we solve for a sharp, noise-free latent image of the night sky by using information from multiple noisy and blurry co-registered exposures of a given region of the sky. In particular, we model the latent image as a \textit{function} of the multiple noisy and blurry exposures, and parametrize this function using a carefully-designed neural network with learnable parameters. By learning the parameters of the network and exploiting the regularizing effect imposed by the network's architecture, \AstroClearNet~yields a latent image which corresponds to a maximum a posteriori (MAP) estimate for a given statistical model of the exposures, as outlined in Section~\ref{ssec:MLE_MAP}.

Our approach is unique in that rather than finding a latent image for a single degraded observation, we combine information from multiple exposures of the same part of the sky in order to generate a single, common latent image. As a result, the neural network architecture proposed in this paper, which is relatively simple and contains few learnable parameters, is suited to processing batches of exposures of the sky, akin to imaging data produced by modern ground-based surveys. 

To demonstrate the effectiveness and performance of \AstroClearNet, we conducted tests using a set of Hyper Suprime-Cam (HSC) exposures, which serve as a precursor to forthcoming imaging data from the Rubin Observatory. The results are highly promising, indicating that the method is well-suited for integration into processing pipelines for studies with ground-based astronomical imaging data.

\section{Modeling the imaging data}
\label{sec:model_astroimaging_data}

\begin{figure*}[htbp]
    \centering
    \includegraphics[width=.99\textwidth]{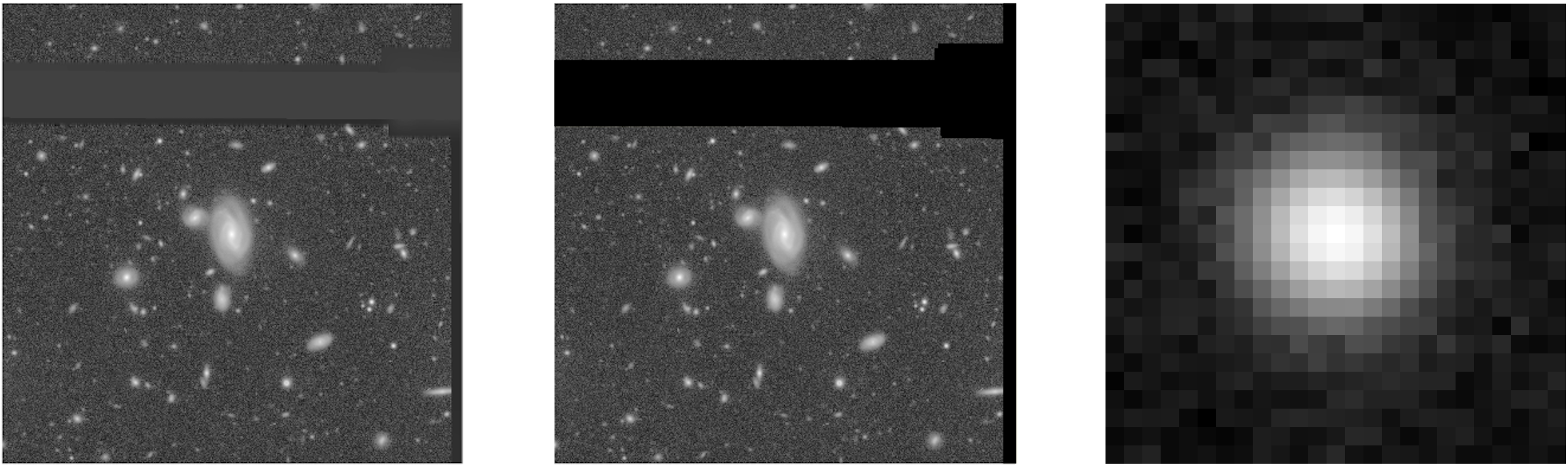}
    \caption{\textbf{Hyper Suprime-Cam (HSC) imaging data.} \emph{Left:} A $1000 \times 1000$ pixel cutout from an HSC exposure, where each pixel $y_i^{(t)}$ is normalized by its standard deviation $\sigma_i^{(t)}$. The exposure contains astronomical sources, including point sources such as stars and extended sources such spiral arm galaxies, which are degraded by noise, blurring due to atmospheric turbulence, and instrument-related artifacts. Prominent artifacts include occlusions due to chip gaps and their borders, which appear as large gray bands with dark gray edges. \emph{Middle:} A binary mask is applied to identify and exclude problematic pixels---such as those affected by chip gaps or sensor edge effects---which are shown in black. These masked regions are disregarded during processing. \emph{Right:} The point spread function (PSF) $f^{(t)}$ associated with this exposure, estimated by the HSC software pipeline, shown on a $25 \times 25$ pixel grid (not to scale). The PSF describes how a point source is spatially blurred in the exposure. Brighter regions indicate stronger diffusion of light, with the PSF wings contributing to extended blurring in the observed image.}
    \label{fig:HSC_data_YMV}
\end{figure*}

We begin by outlining the setup for our method. Modern astronomical surveys produce datasets consisting of calibrated, co-registered, high-resolution exposures of the same region of the sky, which we denote as $y\!=\!\{y^{(1)}, \dots, y^{(n)}\}$. For each \mbox{$t=1,\dots,n$}, the image $y^{(t)}\!\in\!\R^d$ is represented as a $d$-dimensional column vector, corresponding to a noisy, blurry observation captured at time \mbox{$t$}, whose pixel values are denoted as $y^{(t)}_i$ for \mbox{$i = 1,\dots, d$}. To simplify notation, we present the mathematical framework in this paper for images represented as one-dimensional arrays (column vectors). Nonetheless, all models, derivations and algorithms in this manuscript can be readily adapted and have been implemented for two-dimensional image arrays.

We note that the pixel values $y^{(t)}_i$ represent photon counts measured at each pixel in each exposure. Along with these measurements, we also obtain auxiliary data on the variability and usability of each pixel value. In particular, we are given corresponding standard deviations \mbox{$\sigma^{(t)}_i$}, and hence variances \mbox{$v_i^{(t)} \doteq (\sigma^{(t)}_i)^2$}, for these measurements. Additionally, each exposure is accompanied by a mask, denoted as \mbox{$m=\{m^{(1)}, \dots, m^{(n)}\}$}. The masks are binary arrays indicating whether specific pixel values in the exposures are valid measurements. Specifically, for each \mbox{$t \!=\! 1, \dots, n$} and \mbox{$i \!=\! 1,\dots, d$}, the mask entries are defined as follows:
\begin{equation*}
    m^{(t)}_i \doteq
    \begin{cases}
    1, \quad \text{if $y^{(t)}_i$ is an acceptable measurement,} \\
    0, \quad \text{otherwise.}
    \end{cases}
\end{equation*}

Moreover, we are given point-spread functions (PSFs), denoted as \mbox{$f\!=\!\{f^{(1)}, \dots, f^{(n)}\}$}. For each \mbox{$t \!=\! 1, \dots, n$}, the PSF \mbox{$f^{(t)}\!\in \R^{d'}$} is a $d'$-dimensional column vector (where \mbox{$d'< d$}) that represents the convolution kernel (or blur) associated with exposure $y^{(t)}$. These PSFs are typically derived from stars in the exposures, which are identified from a catalog of sources.

We display a concrete example of such imaging data in Figure~\ref{fig:HSC_data_YMV}, which is sourced from the Hyper Suprime-Cam (HSC) survey. This dataset consists of \mbox{$n=33$} co-registered and calibrated $i$-band exposures taken by the HSC telescope, together with their associated masks and PSFs. For more information on the pre-processing of this imaging data, including the registration and calibration of the exposures, as well as PSF modeling, see~\cite{aihara2018hyper,miyazaki2012hyper,bosch2018hyper}.

To model the exposures, we follow the approach proposed in the \emph{ImageMM} framework~\cite{sukurdeep2025imagemm}. Specifically, we model each observed exposure $y^{(t)}$ as the result of convolving the true, background-subtracted, noise-free latent image of the night sky, denoted as $x$, with the corresponding PSF $f^{(t)}$, plus an additive noise term $\eta^{(t)}$. The observed pixel values in each exposure are thus represented by:  
\begin{equation}
    \label{eq:model_exposures}
    y^{(t)}_{i} = ( f^{(t)} \!* x )_{i} + \eta^{(t)}_{i} .
\end{equation}  
The noise terms $\eta^{(t)}_{i}$ are assumed to be \emph{independently} drawn from a probability distribution with a mean of zero and variance $v^{(t)}_{i}$. For instance, one may model the noise terms as \emph{independent}, \emph{mean zero Gaussian random variables} whose variances are given by \mbox{$v_i^{(t)} \doteq ( \sigma^{(t)}_{i} )^2$}, as detailed in~\cite{sukurdeep2025imagemm}. One can also consider noise terms drawn from distributions with heavier tails in order to account for outlier pixel values, as we will explain in the next section. 

Moreover, we highlight that in model~\eqref{eq:model_exposures}, the PSFs and noise terms may vary between exposures, whereas the underlying latent image of the sky remains \emph{common to all exposures}.

\section{The AstroClearNet framework}
\label{sec:AstroClearNet}
To perform multi-frame astronomical image restoration, our goal is to recover the unknown latent image $x$ from model~\eqref{eq:model_exposures} using the exposures and PSFs at hand.

\subsection{MLE and MAP estimation}
\label{ssec:MLE_MAP}
A classical approach for doing so involves solving for the latent image $x$ as a \textit{maximum likelihood estimate} (MLE) of model~\eqref{eq:model_exposures}. To do so, one estimates a latent image $\widehat{x}$ which is most likely to have generated the exposures $y$ and PSFs $f$, by maximizing the joint likelihood \mbox{$p(x \mid y, f)$} of the pixel values of $x$ given the data, i.e., the exposures $y$ and PSFs $f$. Equivalently, the MLE problem entails minimizing the joint negative log-likelihood \mbox{$\mathcal{L}(x \mid y, f) \doteq -\ln p(x \mid y, f)$} of the latent image's pixel values:
\begin{equation}
    \label{eq:deconvolution_mle_general}
    \widehat{x} = \underset{ x \in \mathcal{X} }{\operatorname{argmin}} ~\mathcal{L} \left(x \mid y, f \right).
\end{equation}
This minimization problem is defined over the set of images with non-negative pixel values, denoted as $\mathcal{X} \doteq \{ x \in \R_+^{d+d'-1} \}$. Following~\cite{sukurdeep2025imagemm}, we enforce this non-negativity constraint to obtain physically meaningful maximum likelihood estimates, where sky pixels are zero and source pixels (e.g., stars, galaxies) have positive values. Additionally, the restored image $\widehat{x}$ is padded with \mbox{$d'\!-\!1$} extra pixels to account for flux contributions from sources beyond the telescope’s field of view during the restoration process.

However, MLE techniques often fail to form physically meaningful restorations $\widehat{x}$ of the night sky~\cite{schulz1993multiframe, zhulina2006multiframe, starck2002deconvolution}. One may thus operate under a Bayesian framework, and solve for the latent image as a \textit{maximum a posteriori} (MAP) estimate:
\begin{equation} 
    \label{eq:map_estimate}
    \widehat{x} = \underset{x \in \mathcal{X}}{\operatorname{argmax}} \ \ln p(y, f \mid x) + \ln p(x) .
\end{equation}
In the formulation above, $p(y, f \mid x)$ is the joint conditional distribution of pixel values in the exposures $y$ and PSFs $f$ given those in the latent image $x$, which is given by the joint distribution of the noise terms from~\eqref{eq:model_exposures} (e.g., the Gaussian distribution). Meanwhile, $p(x)$ is a prior distribution on the pixels values of the latent image, for which a handcrafted regularization prior, such as the total variation norm~\cite{ulyanov2018deep}, might typically be used. 

\subsection{AstroClearNet deep image prior}
\label{ssec:AstroClearNet_deep_image_prior}
The key to computing $\widehat{x}$ as a MAP estimate thus lies in an effective choice of regularization prior, which may be challenging in the context of astronomical image restoration. Indeed, the distribution of pixel values in the unknown latent image may be complex, and sometimes indeterminate, especially when working with exposures of unknown regions of the sky. Nevertheless, one can bypass this hurdle and impose an effective prior through the structure of an untrained, generative neural network, i.e., a so-called \textit{deep image prior}~\cite{ulyanov2018deep}. 

Inspired by this approach, we develop a self-supervised multi-frame method for restoring astronomical images, dubbed \AstroClearNet. Our approach is an extension of the flash-no flash method for image-pair restoration proposed in~\cite{ulyanov2018deep}, to the setting of multi-frame image restoration. 

\begin{figure*}[h]
    \centering   
    \includegraphics[width=.95\textwidth, trim={40mm 0mm 40mm 0mm}, clip]{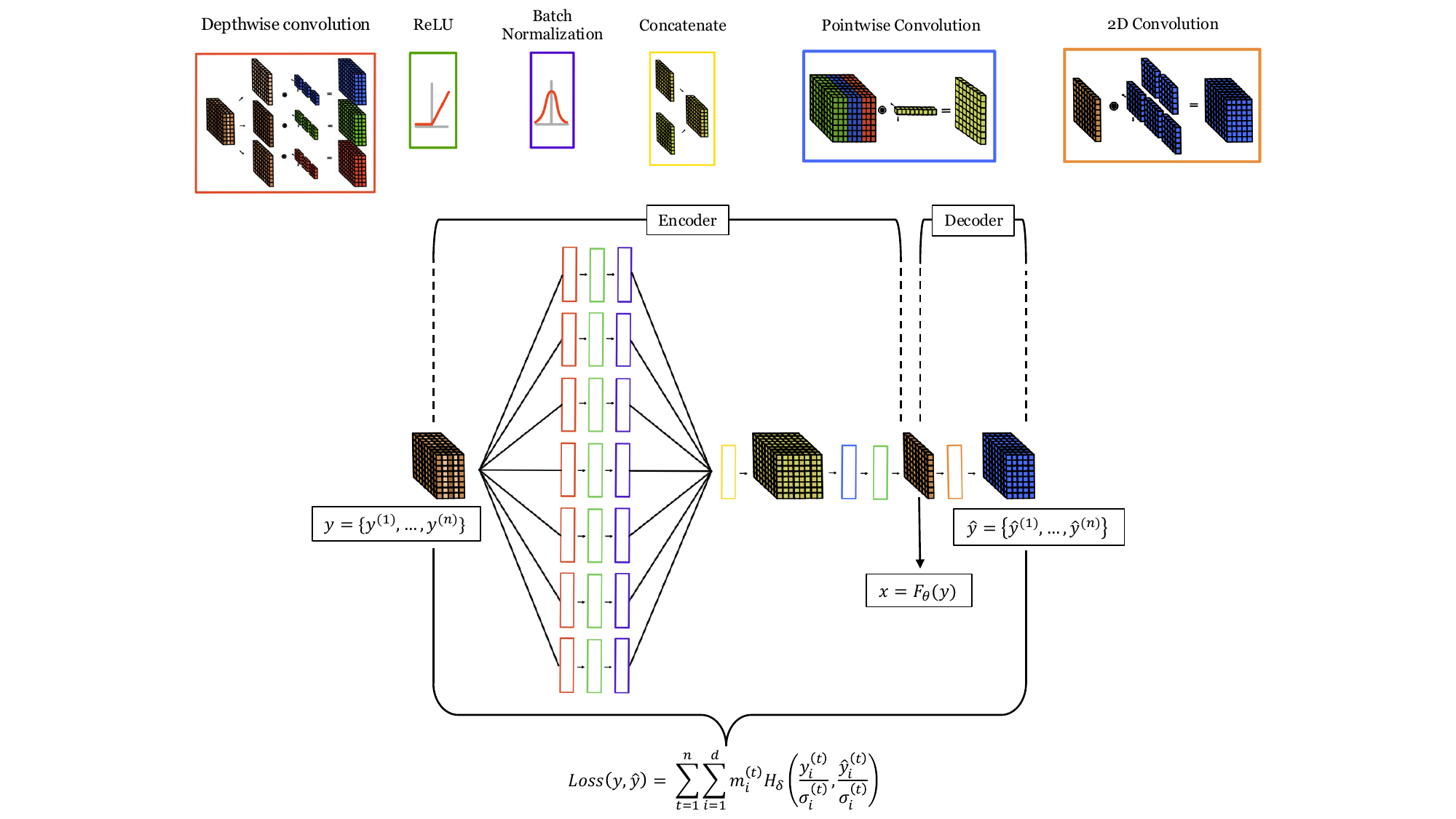}
    \caption{\textbf{Network architecture.} The network takes multiple exposures of the night sky $y = \{y^{(1)},\dots,y^{(n)}\}$ as input. These exposures pass through the ``encoder" of the network, where we extract and combine multi-scale information via various convolutional layers. All input exposures first pass through several depth-wise convolution layers in parallel, before a ReLU activation and batch normalization are applied. The resulting output channels are concatenated, before a pointwise convolution is applied to produce the latent image $x$, which respects desired physical constraints such as non-negativity due to the application of another ReLU activation. We then ``decode" the latent image $x$ via a final 2D convolutional layer to produce the reconstructions $\widehat{y} = \{\widehat{y}^{(1)},\dots,\widehat{y}^{(n)}$\}. This last layer corresponds to the PSFs for each exposure (if these are known), and thus contains fixed, constant weights. Otherwise, the weights in the last layer could be considered as additional learnable parameters, allowing for the extension of \AstroClearNet~to the setting of blind multi-frame image restoration, as outlined in~\ref{app:blind_deconvolution}.}
    \label{fig:network_architecture}
\end{figure*}

As part of our method, we solve for the latent image in a self-supervised manner using a neural network with an encoder-decoder (i.e., ``hourglass") architecture, as illustrated in Figure~\ref{fig:network_architecture}. Similarly to what was outlined in a previous publication by the authors~\cite{navarro2023learning}, the salient feature of \AstroClearNet~consists of encoding the latent image $x$ as a \textit{function} of the multiple exposures \mbox{$y=\{y^{(1)}, y^{(2)}, \dots, y^{(n)}\}$}. We parametrize this function using a neural network $F_{\theta}$ with learnable parameters $\theta$, whose architectural details are outlined in Section~\ref{ssec:AstroClearNet_architecture}. We then decode the latent image \mbox{$x = F_{\theta}(y)$} by convolving it with $n$ convolutional filters \mbox{$f\!=\! \{f^{(1)}, \dots, f^{(n)}\}$} in order to produce reconstructions of our input exposures, which we denote by \mbox{$\widehat{y} = \{\widehat{y}^{(1)}, \dots, \widehat{y}^{(n)}\}$}, where \mbox{$\widehat{y}^{(t)} = f^{(t)} * F_{\theta}(y)$} for each \mbox{$t=1,\dots,n$}. The convolutional filters $f$ correspond to the PSFs for each exposure, if these are known. Otherwise, they could be included as additional learnable parameters of the hourglass network, as outlined in \ref{app:blind_deconvolution}.

\subsection{Training the network}
\label{ssec:training_network}
Under this setup, the task of solving for the latent image $x = F_{\theta}(y)$ boils down to learning the unknown mapping $F_{\theta}$. In turn, this essentially entails tuning the unknown learnable parameters of the network, $\theta$, such that the network generates reconstructions $\widehat{y}$ that are consistent with the input exposures $y$. 

Inspired by the robust restoration framework of~\cite{sukurdeep2025imagemm}, we do so by minimizing the \emph{Huber loss} between our network's inputs $y$ and outputs $\widehat{y}$, namely:
\begin{equation}
    \label{eq:loss_network}
     \theta^* = \underset{\theta}{\operatorname{argmin}} \sum_{t,i} m_i^{(t)} H_{\delta} \left(\frac{y_{i}^{(t)}}{\sigma_{i}^{(t)}}, \frac{\left[ f^{(t)} \!*\! F_\theta(y) \right]_{i}}{\sigma_{i}^{(t)}}\right)
\end{equation}
where $H_{\delta}: \R \times \R \to \R$ is defined as follows:
\begin{equation}
    \label{eq:huber_loss}
    H_{\delta}(y, \Hat{y}) \coloneqq \begin{cases}
        \frac{1}{2} \left(y - \widehat{y} \right)^2  & \text{for } | y - \hat{y} | \leq \delta , \\
        \delta \left( | y - \widehat{y} | - \frac{1}{2} \delta \right)  & \text{otherwise. }
    \end{cases}
\end{equation}
We note that $H_{\delta}$ is applied pixel-wise across all pairs of corresponding pixels in the input exposures $y$ and their reconstructions $\widehat{y}$, which are scaled by their standard deviations. We adopt the Huber loss as it balances sensitivity to small discrepancies and robustness to outliers. Indeed, when the difference between pixel values in $y$ and $\widehat{y}$ is small (i.e., below a threshold $\delta \in \R$), the Huber loss behaves like the mean squared error, which corresponds to the negative log-likelihood function~\eqref{eq:deconvolution_mle_general} of the latent image's pixels values under the assumption of Gaussian noise in model~\eqref{eq:model_exposures}. However, for larger discrepancies exceeding $\delta$, which are typically caused by outlier pixels, the Huber loss transitions to a linear function. This reduces the influence of such outliers on the overall loss, especially compared to the mean squared error. The Huber loss is thus less sensitive to outliers, enabling the recovery of latent images $\hat{x}$ that are robust to the adverse impact of heavy-tailed noise in the exposures. 

For the sake of clarity, we point out that once the network has been trained by finding the optimal network parameter values $\theta^*$ in~\eqref{eq:loss_network}, the estimated latent image $\widehat{x}$ is then computed via a forward pass through the trained encoder of our hourglass network, namely:
\begin{equation}
    \label{eq:latent_image_forward_pass}
    \widehat{x} = F_{\theta^*}(y). 
\end{equation}

\emph{Remark}. On a practical note, the network was implemented in \texttt{TensorFlow}~\cite{tensorflow2015-whitepaper}, and trained using the Adam optimizer~\cite{kingma2014adam} with an initial learning rate of $0.002$, decayed by a factor of $0.95$ every $1000$ steps. We used \texttt{TensorFlow}’s default weight initialization strategy, namely the Glorot uniform initializer~\cite{glorot2010understanding}. For the results reported in Section~\ref{sec:results}, the network was trained for $450$ epochs, and all computations for this paper were performed on the SciServer platform~\cite{taghizadeh2020sciserver} using a compute
engine with an Intel Xeon Gold 6226, 12-core, 2.70GHz
CPU, with a Tesla V100-SXM2 GPU.

\subsection{AstroClearNet network architecture}
\label{ssec:AstroClearNet_architecture}
We now elaborate on the key architectural features of the \AstroClearNet~neural network, which is illustrated in detail in Figure~\ref{fig:network_architecture}. 

\medskip

\noindent \textbf{$\bullet$ Hourglass structure:} An hourglass or encoder-decoder architecture naturally fits our image restoration problem. Indeed, since our goal essentially entails solving an inverse problem to recover the latent image $x$ from the observed exposures based on model~\eqref{eq:model_exposures}, we must first map the input exposures into a latent space to learn $x$, thus requiring an encoder. The decoder then reconstructs the exposures from the latent image, forming a self-supervised learning framework. This approach is essential, as no labeled training data is available for our task. 

\medskip

\noindent \textbf{$\bullet$ Multi-frame method and multi-scale feature maps:} 
Moreover, to effectively learn the latent image, we must combine information from all $n$ observations. This motivates the use of depthwise convolutions in the encoder. Indeed, unlike traditional convolutions, which aggregate information across input channels, depthwise convolutions apply separate kernels to each channel, thus preserving information specific to each observation. Furthermore, this allows us to capture features at multiple scales by applying depthwise convolutions in parallel with different kernel sizes, promoting the learning of features at varying resolutions. Once the input exposures pass through these depthwise convolution layers, the resulting feature maps are merged using pointwise convolutions, thereby generating a single latent image which combines multi-scale information from all exposures. 

\medskip

\noindent \textbf{$\bullet$ Incorporating physical model and constraints:}  
Another key requirement when designing \AstroClearNet~entails ensuring that the network's output $\widehat{y}$ adheres to the physical model of the exposures given in~\eqref{eq:model_exposures}. This principle shapes the decoder’s structure, ensuring that its final layer (a 2D convolution) produces outputs that follow $\widehat{y}^{(t)} = f^{(t)} * x = f^{(t)} * F_{\theta}(y)$ for each $t=1,\dots,n$. By embedding this model directly into the neural network's architecture, we maintain consistency with the underlying physics. Additionally, to enforce the non-negativity constraint on the latent image $x$, we apply a ReLU activation in the last layer of our encoder, thus ensuring the network produces a physically meaningful latent image. 

\medskip

We note that the aforementioned architectural features of \AstroClearNet\ can be interpreted through the lens of MAP estimation outlined in~\eqref{eq:map_estimate}.

Firstly, the encoder $F_{\theta}$, which maps a stack of noisy, blurry exposures to the latent image, serves as an implicit prior on the solution space of latent images---analogous to the role played by deep image priors in related restoration tasks~\cite{ulyanov2018deep}. Indeed, in the absence of training data and without an explicit prior distribution $p(x)$ for the pixel values of the latent image, the architecture of the encoder itself (via the use of convolutional layers, ReLU activations, and multi-scale feature extraction) imposes inductive biases that promote the learning of meaningful latent images. These architectural features essentially constrain the solution space, and guide the optimization of the network's parameters towards the production of plausible image restorations during training, thereby functioning as a parametric prior within the MAP estimation framework.

The decoder, on the other hand, is not learned. It represents a fixed forward model defined by a known physical image degradation process, i.e., atmospheric blurring modeled by convolution with the PSFs, and thus corresponds to the likelihood term $ p(y,f \mid x) $ in the MAP formulation~\eqref{eq:map_estimate}. The use of a fixed, non-learned decoder has the additional effect of further constraining the solution space for the latent image learned by the network, thus further guiding the network towards plausible, physically meaningful latent images.

This clear separation between learned prior and fixed likelihood allows \AstroClearNet\ to maintain physical consistency with the data while recovering meaningful restorations that adhere to the observational model, as we will demonstrate with the results in Section~\ref{sec:results}.

\emph{Remark}. We note that certain aspects of the network architecture and parameter choices in \AstroClearNet, such as the number of convolutional layers and size of the convolutional kernels in the encoder, are guided primarily by empirical considerations. While not derived from first principles, these design decisions reflect heuristics that have proven effective in unsupervised image restoration settings, such as those explored in the deep image prior framework of Ulyanov et al.~\cite{ulyanov2018deep}.

\section{Results and discussion}
\label{sec:results}
\begin{figure*}[h!]
    \centering
    \includegraphics[width=0.325\linewidth, trim={38mm 50mm 53mm 45mm}, clip]{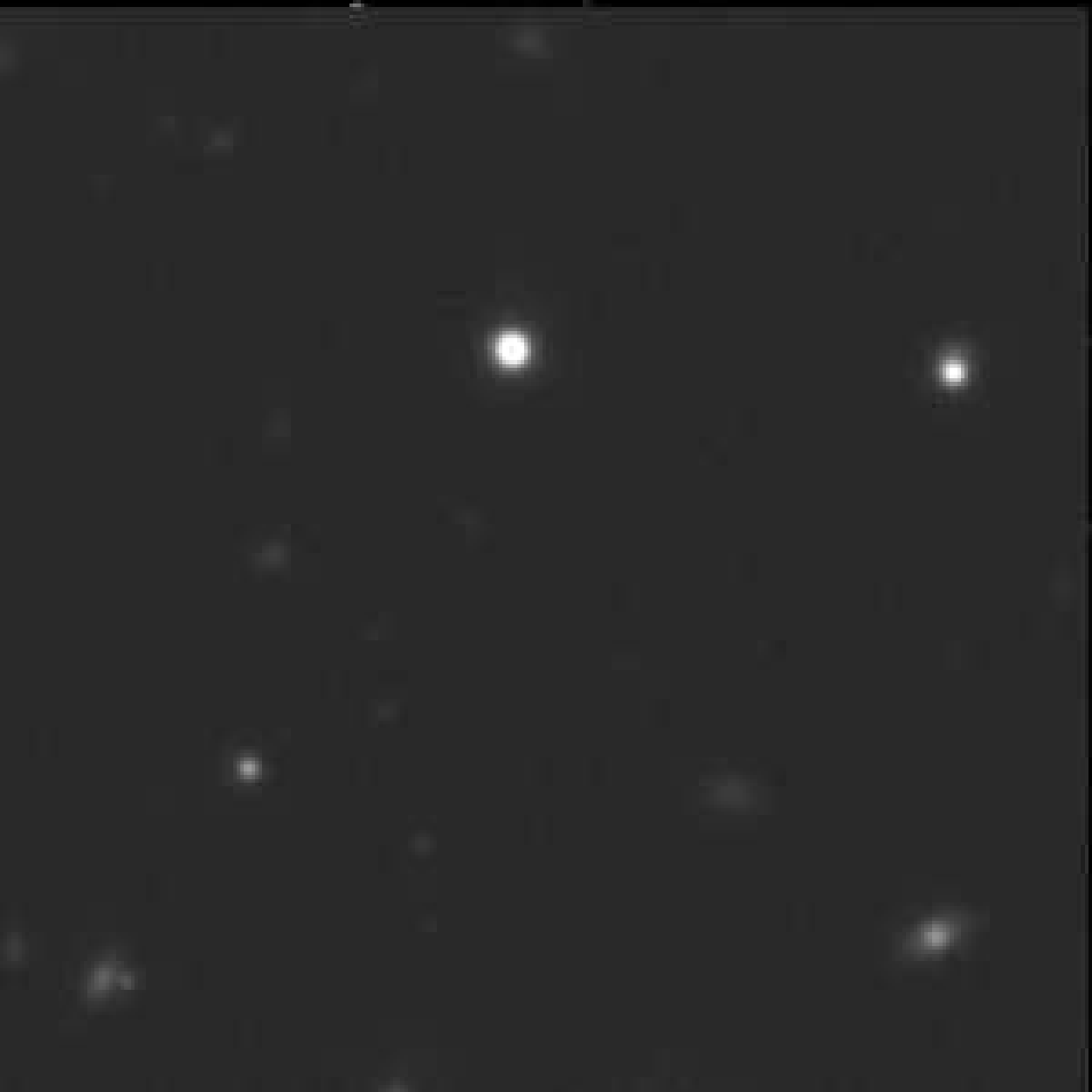}
    \includegraphics[width=0.325\linewidth, trim={38mm 50mm 53mm 45mm}, clip]{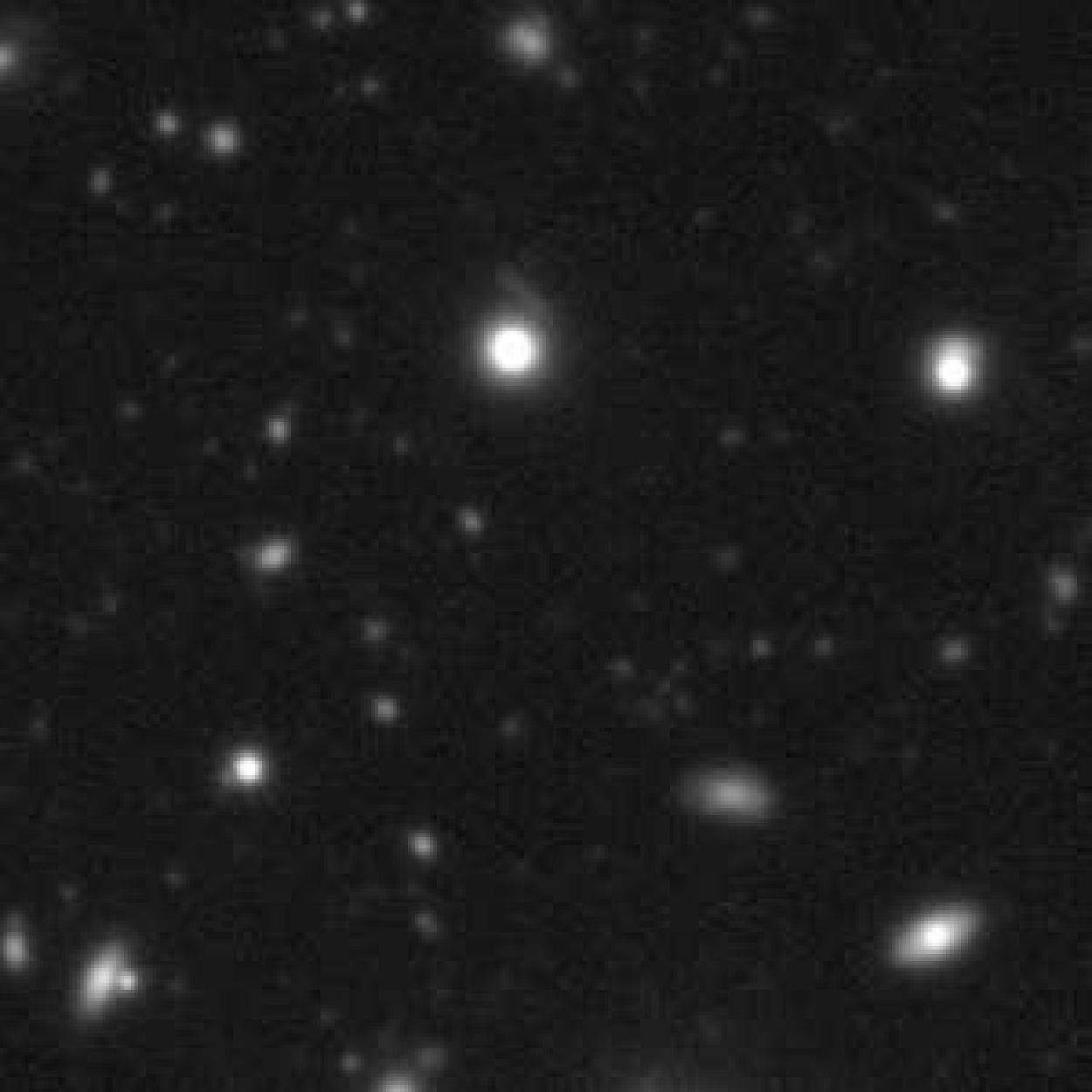}
    \includegraphics[width=0.325\linewidth, trim={38mm 50mm 53mm 45mm}, clip]{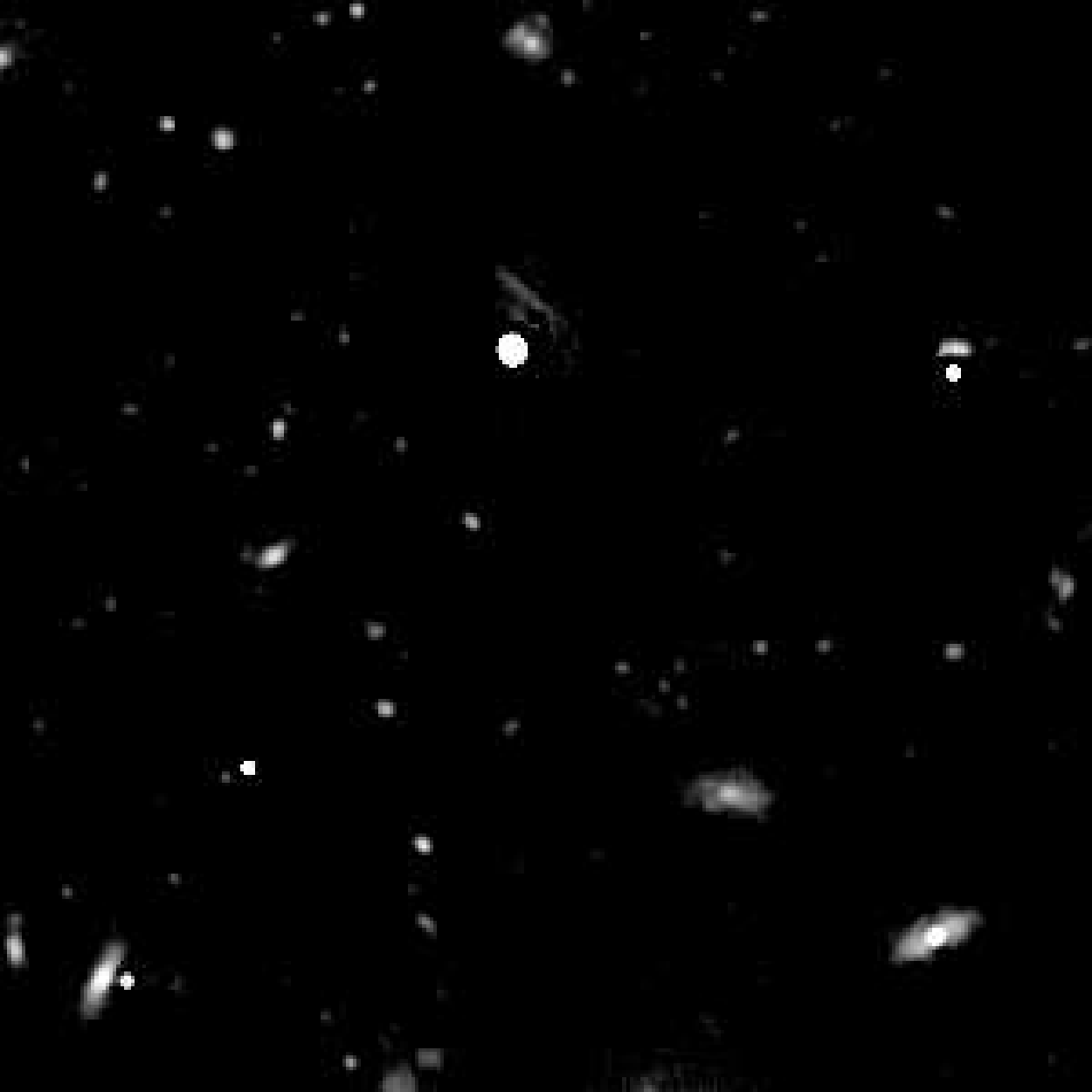}
    \includegraphics[width=0.325\linewidth, trim={5mm 23mm 70mm 60mm}, clip]{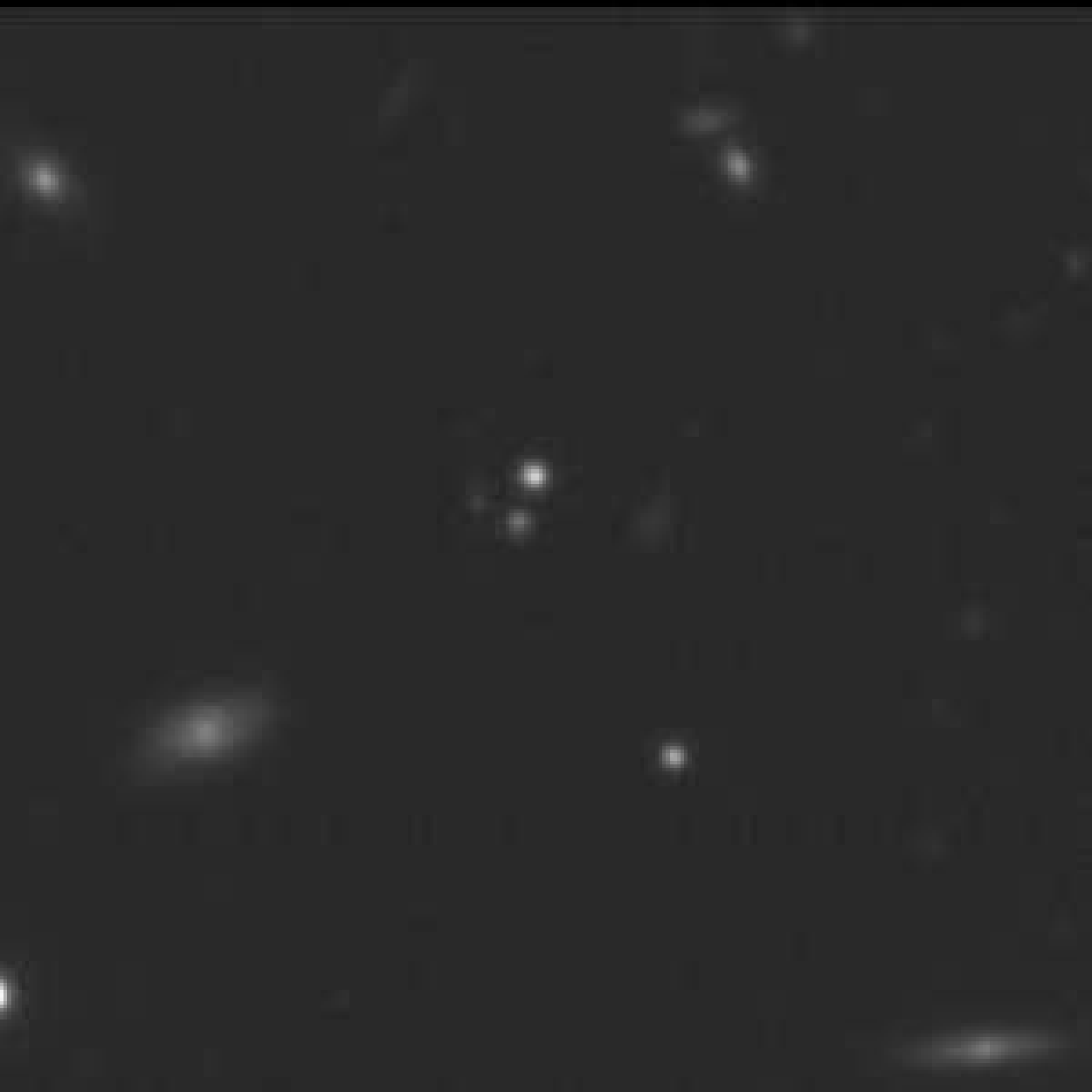}
    \includegraphics[width=0.325\linewidth, trim={5mm 23mm 70mm 60mm}, clip]{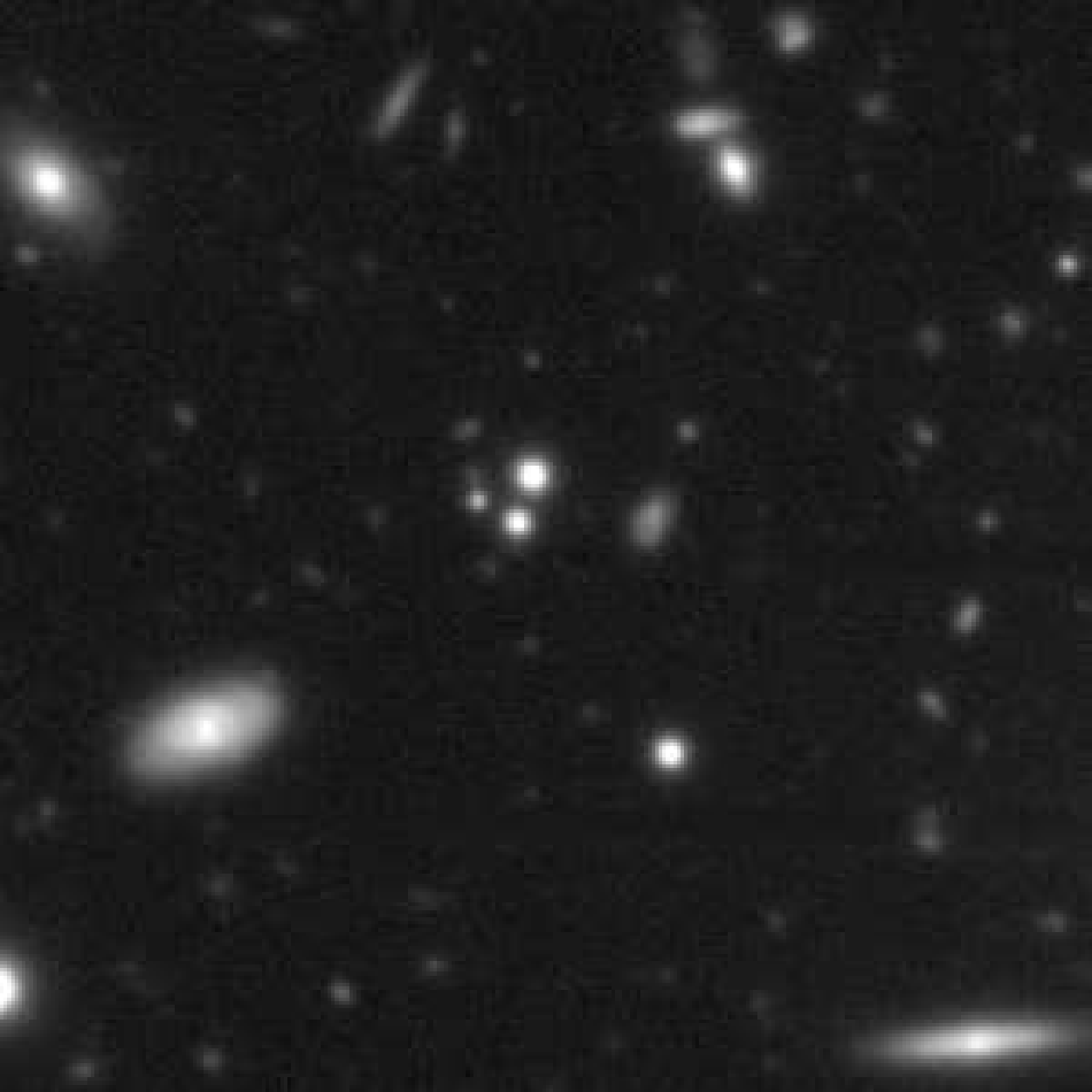}
    \includegraphics[width=0.325\linewidth, trim={5mm 23mm 70mm 60mm}, clip]{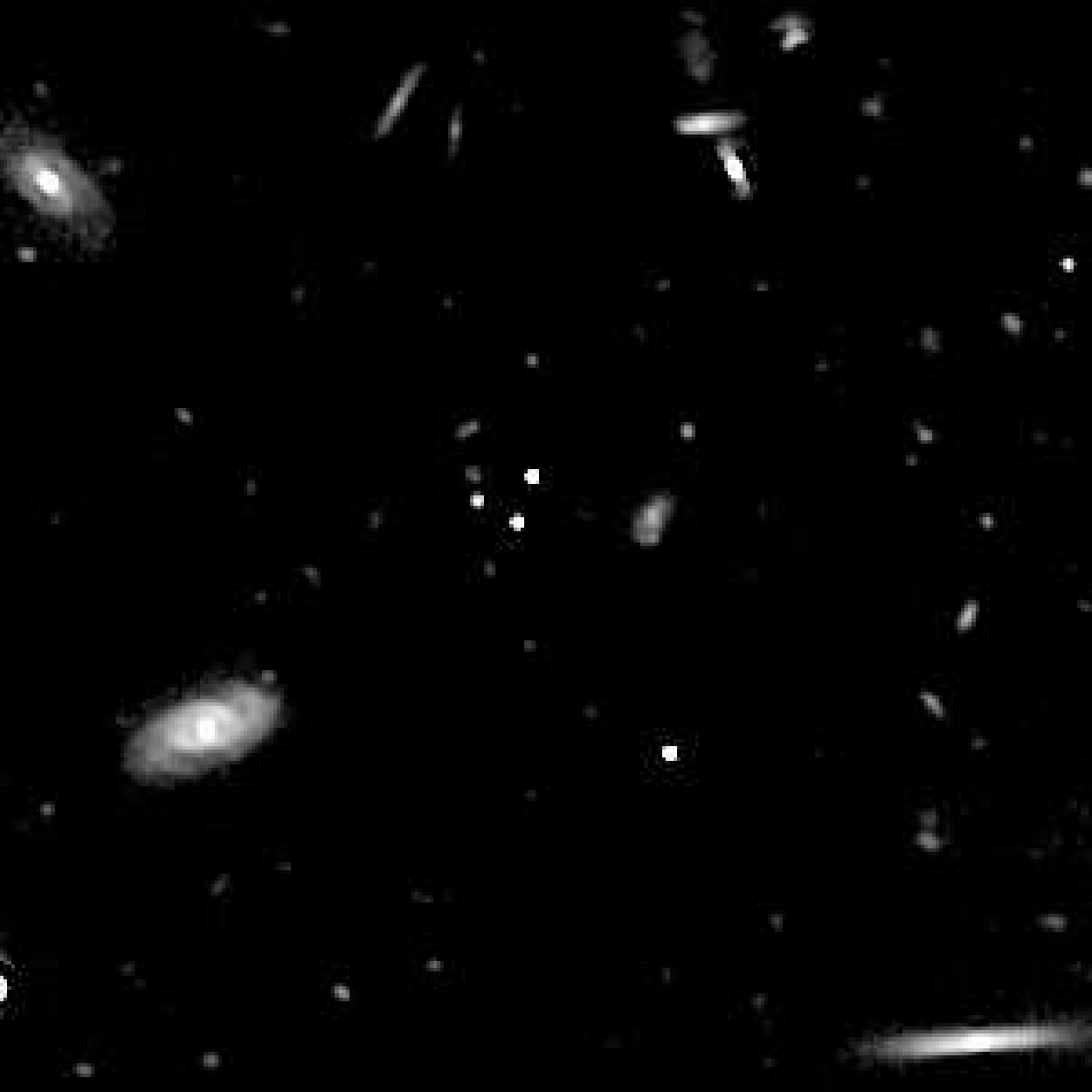}
    \includegraphics[width=0.325\linewidth, trim={3mm 5mm 87mm 88mm}, clip]{figures/exposure1.png}
    \includegraphics[width=0.325\linewidth, trim={3mm 5mm 87mm 88mm}, clip]{figures/coadd1.png}
    \includegraphics[width=0.325\linewidth, trim={3mm 5mm 87mm 88mm}, clip]{figures/xhat1.png}
    \caption[width=\linewidth]{\textbf{Results with HSC imaging data.} \emph{Left:} Cutouts from an HSC exposure, where the sky-background is noisy, and blurry sources are hard to detect, especially when they are small and faint (top row). \emph{Middle:} Corresponding cutouts from the HSC pipeline coadd, in which sky-background noise is reduced, and sources become more distinguishable despite retaining some blur. \emph{Right:} Corresponding cutouts from the latent image $\widehat{x}$ obtained using \AstroClearNet. The estimate $\widehat{x}$ exhibits minimal sky-background noise, and sources appear markedly sharper than those in the coadd, enabling more detailed visualization of fine spatial features such as the shapes and sizes of galaxies (middle and bottom row). All cutouts have been taken from the same field of view of size $4200 \times 4200$ pixels, where each pixel has a spatial scale of 0.168 arcseconds~\cite{aihara2018hyper}.}
    \label{fig:HSC_results}
\end{figure*}

We now present preliminary results obtained by applying \AstroClearNet~on a multi-frame restoration task using ground-based astronomical imaging data from the Hyper Suprime-Cam (HSC) survey, which was described in Figure~\ref{fig:HSC_data_YMV}. Specifically, we used $n=33$ HSC exposures of size $4200 \times 4200$ pixels, together with their corresponding masks, standard deviations and PSFs, in order to recover a latent image of the night sky. Our results are presented in Figure~\ref{fig:HSC_results}.

\subsection{Analysis of preliminary results}
\label{ssec:analysis_results}
Strikingly, Figure~\ref{fig:HSC_results} shows that \AstroClearNet~produces latent images with minimal sky-background noise and noticeably sharper bright sources compared to those in the HSC pipeline coadd. This enhanced clarity reveals fine spatial features---such as the shapes and structures of galaxies---in remarkable detail. We also observe that the method is able to effectively reduce the blur around various types of sources (of varying shapes and sizes), even when sources are not well-separated in the exposures.

To quantitatively assess these visually discernible improvements in \AstroClearNet's restoration relative to the pipeline coadd, we report metrics highlighting the increase in global sharpness and reduction in sky-background noise levels, which are outlined in Table~\ref{tab:sharpness}.

Specifically, we compute a Fourier-based sharpness metric, denoted $S_F$, that evaluates the high-frequency content of each image. We calculate $S_F$ by taking the logarithm of the magnitude of the image's two-dimensional Fourier transform, before taking its mean across the frequency spectrum~\cite{krotkov1988active}. Higher values of the sharpness metric $S_F$ indicate more high-frequency content, likely due to the presence of fine-scale structures and edge detail, and thus greater sharpness in the image. While this metric captures global sharpness effectively, it may also be sensitive to high-frequency noise. We therefore also report the residual sky-background noise levels $\sigma_{\operatorname{sky}}$ in Table~\ref{tab:sharpness}, which is computed as the standard deviation of the background pixels. This calculation was performed using the \texttt{sep} package---the \texttt{Python} implementation of the source extraction software \texttt{SExtractor}~\citep{bertin1996sextractor, barbary2016sep}.

\begin{table}[htbp]
\centering
\small
\begin{tabular}{lcc}
\toprule
  & \textbf{Coadd} & \textbf{\AstroClearNet} \\
\midrule
Sharpness, $S_F$ & 6.084 & 7.007 \\
Noise, $\sigma_{\operatorname{sky}}$ & 3.057 & 0.874  \\
\bottomrule
\end{tabular}
\caption{Quantitative comparison of sharpness and sky-background noise levels between the pipeline coadd and \AstroClearNet's latent images, computed over a field of view of size $4200 \times 4200$ pixels from the HSC survey. Higher values of $S_F$ indicate sharper images, and lower values of $\sigma_{\operatorname{sky}}$ correspond to lower background noise levels.}
\label{tab:sharpness}
\end{table}

The results in Table~\ref{tab:sharpness} show that \AstroClearNet\ produces images with a $15\%$ increase in sharpness compared to the coadd, as measured by the Fourier-based metric \( S_F \), alongside a substantial $71\%$ reduction in background noise, quantified by the sky standard deviation \( \sigma_{\operatorname{sky}} \). These improvements suggest that \AstroClearNet\ enhances fine spatial detail while also suppressing noise, thereby producing sharper and cleaner astronomical images, thus corroborating our visual observations from Figure~\ref{fig:HSC_results}. All reported measurements in Table~\ref{tab:sharpness} were computed over the full field of view from which the cutouts in Figure~\ref{fig:HSC_results} were extracted, ensuring that the comparisons reflect global image properties rather than localized artifacts.

Furthermore, to quantitatively evaluate the fidelity of restorations produced by \AstroClearNet, we compute two widely-used image quality metrics: the peak signal-to-noise ratio (PSNR)~\cite{hore2010image} and the structural similarity index measure (SSIM)~\cite{wang2004image}. Specifically, we compute the average PSNR and SSIM values between \AstroClearNet's restoration and each of the HSC exposures, as well as the PSNR and SSIM values between \AstroClearNet's latent image and the HSC pipeline coadd. These metrics are reported in Table~\ref{tab:ssim_psnr}.

\begin{table}[htbp]
\centering
\small
\begin{tabular}{lcc}
\toprule
 & \multicolumn{2}{c}{\textbf{\AstroClearNet}} \\
\textbf{Metric} & \textbf{vs. Exposures} & \textbf{vs. Coadd} \\
\midrule
PSNR (dB) & 15.78 & 18.88 \\
SSIM  & 0.025 & 0.056 \\
\bottomrule
\end{tabular}
\caption{Comparison of PSNR and SSIM values between \AstroClearNet's restoration versus the HSC exposures, and versus the pipeline coadd, computed across a $4200 \times 4200$ pixel field of view from the HSC survey. Higher PSNR values indicate closer numerical resemblance between the images, and higher SSIM values suggest greater perceptual fidelity between the images.}
\label{tab:ssim_psnr}
\end{table}

The results in Table~\ref{tab:ssim_psnr} show that \AstroClearNet's restorations achieve notably higher PSNR and SSIM values when compared to the pipeline coadd, relative to when they are compared with the original HSC exposures. Specifically, the PSNR improves from $15.78$ dB (vs. exposures) to $18.88$ dB (vs. coadd), corresponding to a relative increase of approximately $19.6\%$. Similarly, the SSIM improves from $0.025$ (vs. exposures) to $0.056$ (vs. coadd), which represents more than a twofold increase in structural similarity.

These findings, when interpreted alongside the improvements in global sharpness and the substantial reduction in background noise reported in Table~\ref{tab:sharpness}, provide further indication that \AstroClearNet\ effectively suppresses noise while enhancing fine-scale features---thereby producing restorations that are both perceptually sharper, while still remaining faithful to the underlying astronomical structures present in the imaging data.

\subsection{Limitations}
\label{ssec:limitations}
Despite the overall effectiveness of \AstroClearNet\ and the encouraging results reported so far, several limitations remain, which we now outline.

\begin{figure*}[h!]
    \centering
    \includegraphics[width=0.99\linewidth]{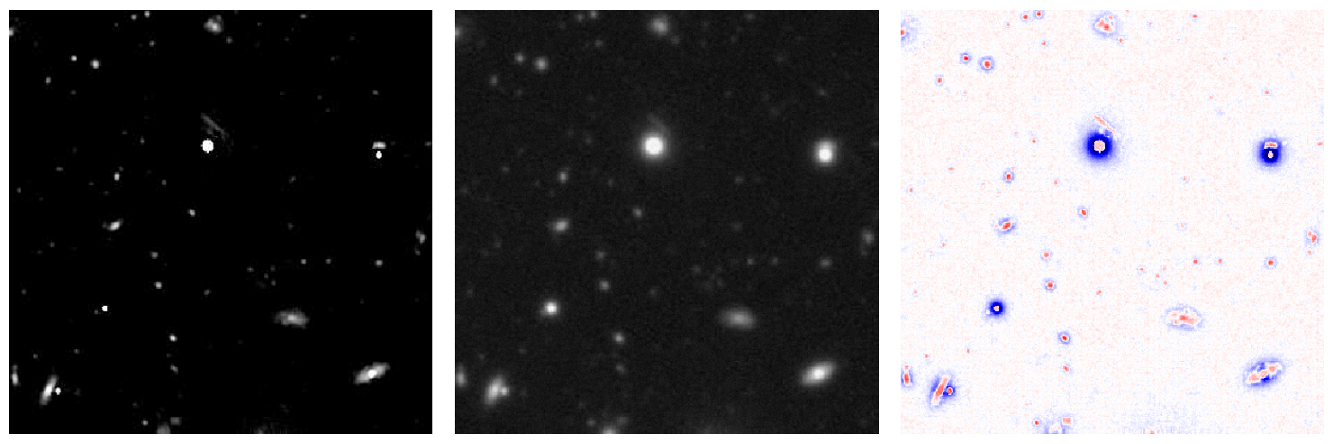}
    \includegraphics[width=0.99\linewidth]{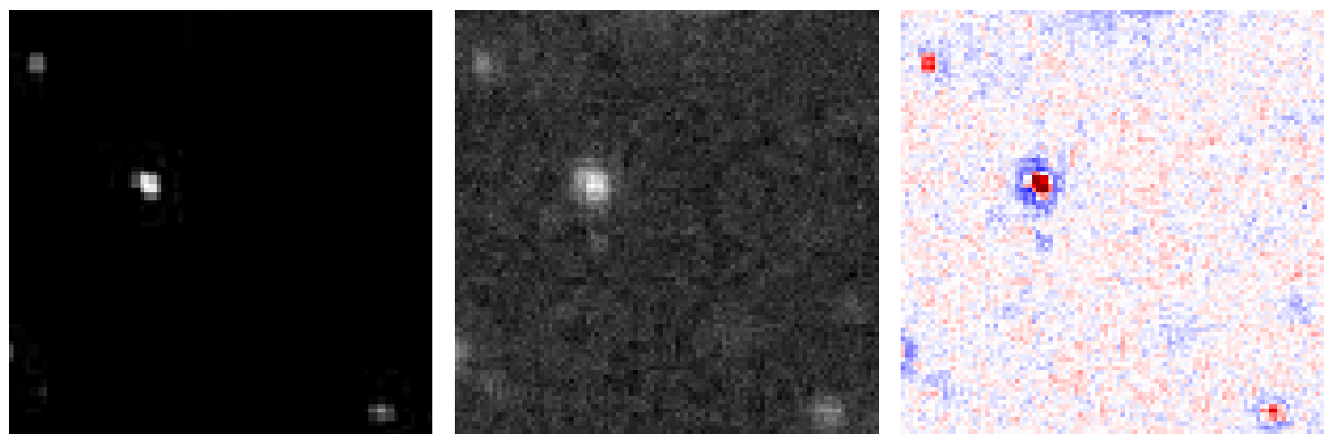}
    \caption[width=\linewidth]{\textbf{Residual between \AstroClearNet's restoration and the HSC pipeline coadd.} \emph{Left:} Cutouts from \AstroClearNet's restoration $\widehat{x}$. \emph{Middle:} Corresponding cutouts from the HSC pipeline coadd. For display purposes, the grayscale values of the coadd cutouts have been chosen to match the range of pixel values the corresponding cutouts of $\widehat{x}$. \emph{Right:} Corresponding cutouts from the residual image, which is computed as the pixel-wise difference between $\widehat{x}$ and the coadd. The colormap for the residual image has been chosen so that \emph{blue} represents negative values, \emph{white} represent a value of zero, and \emph{red} represents positive values, with higher opacity representing pixel intensities with higher absolute values. Blue regions in the residual image therefore represent regions where the latent image $\widehat{x}$ has lower pixel intensities compared to the coadd. Meanwhile, pixels colored in red correspond to regions where $\widehat{x}$ has higher pixel intensities compared to the coadd. \emph{Top row:} A $400 \times 400$ pixel cutout from the telescope's field of view, which contains a wide variety of sources. \emph{Bottom row:} A $100 \times 100$ cutout, where we zoom into a faint sky region.}
    \label{fig:HSC_residuals}
\end{figure*}

\medskip

\noindent \textbf{$\bullet$ Photometric analysis in faint sky regions:} While \AstroClearNet\ improves image quality in relation to coadds and raw exposures, as highlighted in Section~\ref{ssec:analysis_results}, its restorations might not yet be suitable for accurate, fine-grained photometric analysis, especially in faint sky regions. To highlight this, we display cutouts from the residual image between \AstroClearNet's restoration and the HSC pipeline coadd, which is computed as the pixel-wise difference between these respective images, see Figure~\ref{fig:HSC_residuals}.

First, we observe that \AstroClearNet\ tends to produce restorations in which sources appear systematically brighter than in the coadd. This effect is evident in the residual images, where source regions often appear red, indicating a positive flux difference. Closer inspection reveals that \AstroClearNet\ tends to concentrate flux from the surrounding sky background into the sources themselves, particularly near bright objects, hence explaining the blue regions around sources in the residual image. This behavior highlights the need for caution when performing photometric measurements on \AstroClearNet's restorations. Indeed, while \AstroClearNet\ may be preserving global flux levels across the field of view, appropriate calibration procedures must nevertheless be applied to ensure reliable measurement of source properties, such as flux estimates.

Additionally, the bottom row of Figure~\ref{fig:HSC_residuals}, which depicts a field of view from a faint sky region, suggests that \AstroClearNet\ may be less reliable in detecting small, low-surface-brightness sources compared to advanced restoration methods, such as \emph{ImageMM}~\cite{sukurdeep2025imagemm}. The residual images reveal several faint blue patches in areas that visually appear to correspond to the sky background, potentially indicating the presence of low-brightness sources. These negative residuals imply that \AstroClearNet’s restoration underestimates flux in these regions relative to the coadd, possibly even suppressing such sources. If these sources are indeed real, they may be more easily detectable in the coadd, underscoring a limitation of \AstroClearNet\ in recovering faint structures in low signal-to-noise regimes.

One possible explanation for the suppression of faint sources in \AstroClearNet’s restorations is the interaction between ReLU activations and background subtraction errors in the input exposures. Since the ReLU nonlinearity enforces non-negativity, any residual negative background fluctuations in the exposures---potentially introduced during imperfect background subtraction---are effectively clipped to zero, thus systematically suppressing low-signal regions and obscuring faint sources. A second contributing factor may be pixel correlations introduced during the registration of the exposures. Indeed, interpolation steps can create spatial dependencies that are not explicitly modeled in the framework, potentially affecting the fidelity of faint structures in the latent image.

Further comprehensive photometric evaluations and analysis are thus necessary to rigorously assess and understand the detection limits attainable through \AstroClearNet’s restorations.

\medskip

\noindent \textbf{$\bullet$ Generalization beyond the HSC imaging data:} While the results presented in this study demonstrate the promise of \AstroClearNet\ on HSC data, a key limitation is that the method has not yet been validated across data from other telescopes or instruments. The performance of the framework under different optical systems, sensor configurations, and observing conditions---such as spatially-varying point spread functions, different noise characteristics, and varying pixel scales—remains to be systematically investigated. Although the general formulation of \AstroClearNet\ is not specific to HSC data and is intended to be applicable to a wide range of multi-frame imaging scenarios, we acknowledge that its robustness and adaptability beyond the current test case must be established through future empirical studies.

\medskip

\noindent \textbf{$\bullet$ Computational limitations:} On a computational level, one of \AstroClearNet's potential shortcomings is the network’s possible sensitivity to its initial weight configuration, as the optimization of its learnable parameters may vary depending on their initialization scheme. In this study, we adopt TensorFlow’s default weight initialization strategy~\cite{glorot2010understanding}, but acknowledge that this may introduce variability in the results. Furthermore, memory limitations during training---which are bound to arise when handling high-dimensional astronomical imaging data---also restrict the batch size, which can in turn influence the stability and efficiency of convergence. Future work should aim to systematically evaluate the reproducibility and robustness of the framework under various initialization schemes and hardware environments.

\subsection{Future work}
\label{ssec:future_work}
The limitations discussed so far suggest several paths for improving on the preliminary results obtained with \AstroClearNet. We outline them below.

\medskip

\noindent \textbf{$\bullet$ Super-resolution:} For instance, it may be possible to enhance the latent images by performing multi-frame image restoration in a higher spatial resolution, as shown in~\cite{lee2017robust, sukurdeep2025imagemm}. This would entail modifications of \AstroClearNet's network architecture to allow for learning a super-resolved latent image $x$ with sub-pixel detail in galaxies and stars, which may facilitate the detection of small, faint sources. 

\medskip

\noindent \textbf{$\bullet$ Uncertainty quantification:} Moreover, an interesting avenue for future work involves leveraging the differentiability of the network (with respect to the exposures), i.e., the existence of $\partial_y F_{\theta}(y)$, in order to propagate uncertainties from the input exposures through the restoration process. This would allow us to construct confidence intervals for the restored latent images, providing valuable information for subsequent scientific analyses. 

\medskip

\noindent \textbf{$\bullet$ Comprehensive quantitative assessment and benchmarking:} More broadly, a key avenue for the future development of \AstroClearNet\ involves performing a comprehensive quantitative evaluation of its restoration performance. This could be pursued by comparing \AstroClearNet's latent images against restorations produced by other state-of-the-art methods, using standard image quality metrics, such as the PSNR and SSIM metrics. Alternatively, with an emphasis on photometric accuracy, one could compare source fluxes or other key photometric quantities measured from \AstroClearNet's outputs to those in established astronomical catalogs. Although we considered including such evaluations in the present study, we ultimately chose not to, as the current results do not yet match the fidelity of more established restoration methods, as shown in Section~\ref{ssec:limitations}. Nevertheless, as \AstroClearNet\ continues to mature, we intend to undertake rigorous benchmarking using both simulated and real datasets. In particular, the authors plan on rigorously comparing \AstroClearNet~with the \emph{ImageMM} method~\cite{sukurdeep2025imagemm}, as well as exploring the possibility of combining the two methods to leverage the strengths of each.

\medskip

\noindent \textbf{$\bullet$ Broader applicability and future scientific utility}: Looking ahead, another important direction involves extending the \AstroClearNet\ framework to datasets beyond the Hyper Suprime-Cam (HSC). As a general-purpose multi-frame image restoration method, \AstroClearNet's ability to handle noisy and blurry images has the potential to benefit a wide range of astronomical imaging modalities. In particular, we aim to evaluate the method on data from upcoming deep optical surveys such as the Rubin Observatory’s Legacy Survey of Space and Time (LSST)~\cite{ivezic2019lsst}, and are also interested in extending the framework to other wavelength regimes, including near-infrared and ultraviolet imaging, where noisy conditions are common and traditional coaddition may not achieve optimal signal-to-noise ratios. These investigations will help assess the generalization performance of \AstroClearNet\ in the presence of diverse instrumental and atmospheric effects, and guide any necessary architectural or training adjustments to improve its adaptability across datasets.

More broadly, improved restorations produced by \AstroClearNet\ could support critical downstream tasks in astronomy, such as the detection of faint sources, morphological classification of galaxies, and accurate photometric measurements---especially in challenging low signal-to-noise regimes. While the current implementation remains preliminary, these directions underscore the long-term scientific utility of unsupervised, data-driven restoration methods in astronomical analysis pipelines.

\appendix
\section{Joint estimation of PSFs}
\label{app:blind_deconvolution}
As mentioned in Section~\ref{ssec:AstroClearNet_deep_image_prior}, one can extend the \AstroClearNet~framework in order to perform multi-frame blind image restoration. 

In this setting, the PSFs $f\!=\!\{f^{(1)}, \dots, f^{(n)}\}$ corresponding to each exposure $y\!=\!\{y^{(1)}, \dots, y^{(n)}\}$ are assumed to be unknown, and thus, one may jointly estimate the latent image $x$ and these PSFs $f$ as maximum a posterior (MAP) estimates of model~\eqref{eq:model_exposures}. This is achieved by computing:
\begin{equation} 
    \label{eq:map_estimate_app}
    \widehat{x}, \widehat{f} = \underset{x, f}{\mathrm{argmax}} \ \ln p(y \mid x, f) + \ln p(x, f) .
\end{equation}
Note that $p(y \mid x, f)$ is the conditional distribution of the pixel values in the exposures $y$ given those in the latent image $x$ and PSFs $f$, which, as outlined in Section~\ref{ssec:MLE_MAP}, is given by the joint distribution of the noise terms from~\eqref{eq:model_exposures} (e.g., the Gaussian distribution). Meanwhile, $p(x, f)$ is a joint prior distribution on the pixels values of the latent image and PSFs, for which a handcrafted regularization prior is typically used. 

Due to the difficulties in coming up with an effective prior, especially in the context of astronomical image restoration, one may therefore compute $\widehat{x}$ and $\widehat{f}$ as MAP estimates by leveraging the \AstroClearNet~deep image prior, exactly as outlined in Section~\ref{ssec:AstroClearNet_deep_image_prior}.

In particular, when training the hourglass network from Figure~\ref{fig:network_architecture}, one would jointly learn the network weights of the encoder, $\theta$, as well as the weights of the decoder, i.e., the final 2D convolution layer $f$:

\begin{equation}
    \label{eq:loss_network_joint}
     \theta^*, f^* = \underset{\theta, f}{\operatorname{argmin}} \sum_{t,i} m_i^{(t)} H_{\delta} \left(\frac{y_{i}^{(t)}}{\sigma_{i}^{(t)}}, \frac{\left[ f^{(t)} \!*\! F_\theta(y) \right]_{i}}{\sigma_{i}^{(t)}}\right)
\end{equation}

The optimal weights of the decoder $f^*$ obtained after training the network would thus correspond to the MAP estimates of the unknown PSFs for each exposure.

\section*{Acknowledgements}
The authors thank Yusra AlSayyad for providing access to the Hyper Suprime-Cam imaging data, and for extending valuable support with regards to the use of the LSST Science Pipelines for processing the data. 

Y.S. and F.N. gratefully acknowledge support from the NVIDIA Academic Hardware Grant Program. 

T.B. gratefully acknowledges support from the National Science Foundation (Award 1909709 and Award 2206341). 

This research makes use of the SciServer science platform (\url{www.sciserver.org}). SciServer is a collaborative research environment for large-scale data-driven science. It is being developed at, and administered by, the Institute for Data Intensive Engineering and Science at Johns Hopkins University. SciServer is funded by the National Science Foundation through the Data Infrastructure Building Blocks (DIBBs) program and others, as well as by the Alfred P. Sloan Foundation and the Gordon and Betty Moore Foundation.

\bibliographystyle{elsarticle-num}
\bibliography{references}

\end{document}